\documentclass[aps,prb,citeautoscript,superscriptaddress,reprint,twocolumn]{revtex4}

\usepackage{graphicx}
\usepackage{amsfonts,amsmath,amssymb,amsthm}
\usepackage{epstopdf}
\usepackage{upgreek,xspace}
\usepackage{chngcntr}
\usepackage[colorlinks,allcolors=blue]{hyperref}
\usepackage[version=3]{mhchem} 

\newcommand{\be}{\begin{equation}}
\newcommand{\ee}{\end{equation}}

\newcommand{\Tc}{$T_\mathrm{c}$}

\begin{document}

\title{Monoclinic LaSb$_2$ Superconducting Thin Films}

\author{Adrian~Llanos}
\affiliation{Department of Applied Physics and Materials Science, California Institute of Technology, Pasadena, California 91125, USA.}
\affiliation{Institute for Quantum Information and Matter, California Institute of Technology, Pasadena, California 91125, USA.}

\author{Giovanna~Campisi}
\affiliation{Department of Physics and W. M. Keck Computational Materials Theory Center, California State University, Northridge, Northridge,
California 91330, USA.}

\author{Veronica~Show}
\affiliation{Department of Applied Physics and Materials Science, California Institute of Technology, Pasadena, California 91125, USA.}
\affiliation{Institute for Quantum Information and Matter, California Institute of Technology, Pasadena, California 91125, USA.}

\author{Jinwoong~Kim}
\affiliation{Department of Physics and W. M. Keck Computational Materials Theory Center, California State University, Northridge, Northridge,
California 91330, USA.}

\author{Reiley Dorrian}
\affiliation{Department of Applied Physics and Materials Science, California Institute of Technology, Pasadena, California 91125, USA.}
\affiliation{Institute for Quantum Information and Matter, California Institute of Technology, Pasadena, California 91125, USA.}

\author{Salva~Salmani-Rezaie}
\affiliation{Department of Materials Science and Engineering, The Ohio State University, Columbus, Ohio 43210, USA}

\author{Nicholas~Kioussis}
\affiliation{Department of Physics and W. M. Keck Computational Materials Theory Center, California State University, Northridge, Northridge,
California 91330, USA.}

\author{Joseph~Falson}
\email{falson@caltech.edu}
\affiliation{Department of Applied Physics and Materials Science, California Institute of Technology, Pasadena, California 91125, USA.}
\affiliation{Institute for Quantum Information and Matter, California Institute of Technology, Pasadena, California 91125, USA.}

\begin{abstract}
Rare-earth diantimondes exhibit coupling between structural and electronic orders which are tunable under pressure and temperature. Here we present the discovery of a new polymorph of LaSb$_2$ stabilized in thin films synthesized using molecular beam epitaxy. Using diffraction, electron microscopy, and first principles calculations we identify a YbSb$_2$-type monoclinic lattice as a yet-uncharacterized stacking configuration. The material hosts superconductivity with a $T_\mathrm{c}$~=~2~K, which is enhanced relative to the bulk ambient phase, and a long superconducting coherence length of 140nm. This result highlights the potential thin film growth has in stabilizing novel stacking configurations in quasi-two dimensional compounds with competing layered structures.
\end{abstract}

\flushbottom
\maketitle

\begin{figure*}[ht]
    \centering
    \includegraphics[width=160mm]{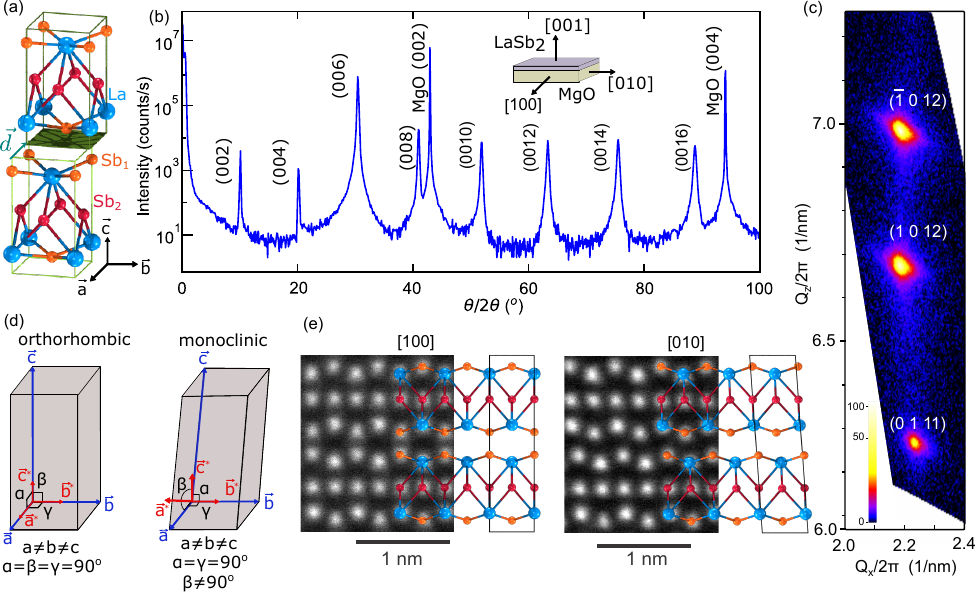}
    \caption{a.) Two  QL building blocks of LaSb$_2$, where the upper QL block is shifted relative to the lower one by lateral displacement(b) $\theta/2\theta$ diffraction pattern from LaSb$_2$. (c) Reciprocal space map taken along the $h\ = 1$ rod. Splitting at finite $Q_x$ indicates a monoclinic tilt. (d) Schematic unit cells of the orthorhombic and monoclinic crystal systems. (e) Atomic arrangements obtained in transmission electron microscopy within domains with an overlay of the calculated unit cell modified with experimental lattice constant values.}
    \label{Fig1}
\end{figure*}

Layered intermetallics featuring the square-net structural motif form a versatile platform\cite{Hoffmanbreak,ironbasedreview, klemenz:2019, yumigetareview:2021} for engineering electronic phases through chemical control over lattice symmetries \cite{Ru:2008chemicalpressure,Ni2008,Lei:2019} and stacking configurations.\cite{suppressionCDW} Using common experimental tuning knobs such as chemical substitution, pressure ($P$), and temperature ($T$), it is known that structural configurations compete and can be modified with a subsequent influence the flavor of electronic ground state of the compound.\cite{dimasi:1996,shin:2005,Huang2008,Xiang2020,singha:2021,weinberger:2023} Antimony-based $Ln$Sb$_2$ (Ln = lanthanide element) materials form stoichiometric crystals and refined structural analysis\cite{LaSb2_bulk} describes the unit cell as being formed by two quintuple layer (QL) blocks, each consisting of two Ln-Sb corrugated layers sandwiching a two-dimensional Sb square net sheet. Variations in electron count on the square net Sb sites as determined by the rare-earth ionization state leads to a variety of bonding and stacking arrangements.\cite{hoffman_hypervalent} In the case of EuSb$_2$, the Eu$^{2+}$ valence state causes a distortion of the Sb square net into zig-zag chains and a monoclinic crystal structure \cite{HULLIGER1978,Ohno2021} while the orthorhombic structure of YbSb$_2$ is stabilized\cite{YbSb2_bulk} due to a small admixture of Yb$^{3+}$ ions within a predominantly 2+ network.\cite{Tremel:1987} Finally, in the case of (La, Ce, Nd, Sm)Sb$_2$ compounds, the 3+ valence again gives rise to an orthorhombic crystal in the so-called SmSb$_2$ structure type.\cite{LaSb2_bulk,Budko:1998}  This latter structure differs from the YbSb$_2$ most notably in the stacking arrangement of QL along the $c$-axis direction, which in this context, can be conceptualized by a displacement vector $\vec{d}$ within the $a$-$b$ plane, as illustrated in Figure \ref{Fig1}a. 

LaSb$_2$ stands out as a particularly sensitive structure which is susceptible to electronic and structural instabilities when tuning $T$ and $P$. Bulk crystals grown with the self-flux method \cite{flux_growth} form large, micaceous crystals with high residual resistivity ratios (RRR) \cite{Budko:1998, fischer2019} and non-saturating positive magnetoresistance (MR).\cite{Budko:1998, young:2003} Pronounced hysteretic electrical resistance when sweeping $T$ is resolved under ambient $P$ conditions.\cite{ditusa:2011, Luccas:2015, weinberger:2023, palacio:2023} Under application of moderate $P$, the hysteretic feature is rapidly suppressed to low $T$ and is completely absent by $P \approx 12$~kbar. Accompanying this suppression is a rapid sharpening of the superconducting (SC) transition under $P$ and an increase in SC critical temperature (\Tc) to a maximum of \Tc~$\approx$~2~K.\cite{Budko:2023, guo:2011, weinberger:2023} However, due to the micacity of crystals, reliable structural analysis at high $P$ remains challenging and therefore complicates a full understanding of correlation between structural and electronic properties. 

Molecular-beam epitaxy (MBE) as a synthesis technique has found much application in the engineering of low-temperature electronic ground states primarily through dimensional confinement\cite{haoFeSe, shiogai2016,jeffrey2011enhancement,falson2020type} and epitaxial strain.\cite{Bozovic:2002,ruf2021strain,engelmann2013strain} Here, we use this technique to synthesize and study LaSb$_2$ thin films on MgO (001) (SG: Fm$\bar{3}$m, a = 4.21\AA). Despite the lattice constants being relaxed in the (001)-oriented films, we discover a not previously observed crystal structure derived from the YbSb$_2$ structure type with an additional monoclinic shear. First principles calculations suggest this layering arrangement as the lowest energy state, even compared to the observed SmSb$_2$ structure type of LaSb$_2$ single crystals commonly observed in bulk crystals. The electronic structure is characterized as a multi-band system with one hole-like band derived from the La-Sb corrugated layer and multiple electron-like bands from the Sb sheet. The films display no resistance anomaly at high $T$ indicative of a structural phase transition and are superconducting with a \Tc~=~2.03~K, which is enhanced relative to the bulk ambient where \Tc~$\approx$~1~K. The ability to stabilize stacking configurations of quasi-2D materials not observed in their bulk counterparts emphasizes the extra degrees of freedom available to thin film synthesis of this broad class of materials.

The structure of $Ln$Sb$_2$ materials is composed of QL blocks which can form a variety of stacking configurations depending on their relative displacement\cite{LaSb2_bulk,YbSb2_bulk}. This is depicted in Figure \ref{Fig1}a with the vector $\vec{d}$ which represents the displacement in the $a$-$b$ plane. Each QL contains one rare earth site, in this case La (blue), and two different Sb sites, noted as Sb$_1$ (orange) and Sb$_2$ (red) in Figure 
\ref{Fig1}a. 
Out-of-plane X-ray diffraction of films presented in Figure \ref{Fig1}b shows a single-phase, $c$-axis oriented film with a lattice parameter of $c = 17.62$\r{A}. This value is compressed relative to the value found in bulk LaSb$_2$ in the SmSb$_2$ structure where $c = 18.56$\r{A}.\cite{LaSb2_bulk} A film of thickness $t$ = 144nm was used for determination of the lattice parameters. Rocking curve analysis reveals a full-width-half max of 0.03$^\circ$ indicating low mosaicity and a phi-scan using the (0 1 9) reflection reveals an [100]$_{LaSb_2}\ ||\ [100]_{MgO}$ epitaxial relationship with additional domains oriented [100]$_{LaSb_2}\ ||\ [110]_{MgO}$ (Figure \ref{suppfig:RC-Phi}).

Asymmetric reciprocal space mapping (RSM) (Figure \ref{Fig1}c) along the $h = 1$ truncation rod was used to determine the in-plane lattice constants. The data reveals a twinned crystal structure with sets of spots corresponding to values from the two different crystal axes. For reflections corresponding to the $a$-axis, a pair of spots ((1 0 12) and ($\bar{1}\ 0\ 12)$) are observed, whereas intensity at (0 1 11) is associated with diffraction from the $b$-axis. To simplify comparison with similar, bulk crystal polymorphs, we adopt a setting where the $c$-axis is always taken to be the long axis. The corresponding $a$ and $b$ axis lattice constants and were found to be $a = 4.52$\r{A} and $b = 4.43$\r{A}. The similarity of the lattice constant values for the grown film ($\frac{a-b}{a}\approx2\%$) and their $\approx6\%$ mismatch with the cubic MgO substrate result in a relaxed film with twinned domains wherein both $a$ and $b$ axis predominantly align with the $<100>$ directions of the substrate. For comparison, the bulk ambient phase adopts the SmSb$_2$ structure (space group No.64 $Cmca$) with $a = b= 4.42\text{\AA}, \gamma = 91.27$.\cite{footnote, LaSb2_bulk}

The distribution of the (1 0 12) and ($\bar{1}\ 0\ 12$) spots along $\vec{Q_z}$ reveals a monoclinic structure for the crystal, which can be understood by referring to Figure \ref{Fig1}d. When a crystal distorts from orthorhombic to monoclinic, the $\vec{a*}$ reciprocal lattice vector forms a finite angle with $\vec{a}$ so as to maintain orthogonality with $\vec{c}$. The reduced symmetry of the monoclinic structure relative to the MgO substrate leads to domains oriented 180$^{\circ}$ with respect to each other and the appearance of both $(h 0 l)$ and $(-h 0 l)$ reflections along the same direction. Thus a single spot corresponding to $(h 0 l)$ in the case of an orthorhombic crystal appears to ``split" along $\vec{Q_z}$ into two spots corresponding to the two rotated domains. The angle between either of these spots and the horizontal can be used to determine the monoclinic tilt which we find to be $\beta$ = 85.95$^{\circ}$.

Observed extinction conditions (Figure \ref{suppfig:RSM-SI}) imply an A-centered monoclinic cell and are consistent with three possible space groups ($A2/m$, $A2$, $Am$).\cite{tables} In Figure \ref{Fig1}e we present a pair of scanning transmission electron microscopy (TEM) images labelled by the zone axis of the crystalline material. An extended set of micrographs is presented in the supplementary information, showing that domains occur in varying proportions randomly in the sample, consistent with the X-ray data. Energy dispersive x-ray spectroscopy taken in the TEM allows for chemical identification of the atoms in the unit cell (Figure \ref{suppfig:EDX}) and confirms the 1:2 La:Sb stoichiometry within an accepted tolerance of $\approx$10\% for EDX in STEM.\cite{lugg:2015,macarthur:2017} Line cuts along Sb layers (Figure \ref{suppfig:linecut}) reveal that Sb sheets form planar layers in contrast to the puckered layers found in the YbSb$_2$ structure.\cite{YbSb2_bulk} Our proposed crystal structure was obtained from density functional theory calculations, the details of which will be discussed in the next section. In Figure \ref{Fig1}(e) we overlay the calculated structure on the TEM images, with lattice parameters taken from experiment, which confirms the proposed atomic arrangements. The space group for the LaSb$_2$ films was determined to be $A2/m$ (No. 12). 

\begin{figure}[ht]
	\centering
t	\includegraphics[width=8.2cm]{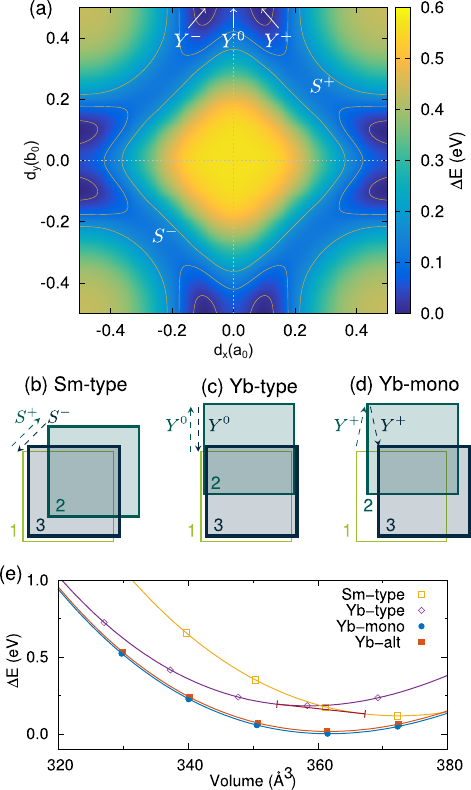}
	\caption{ 
	(a) Energy landscape of sliding two LaSb$_2$ QL building blocks in Figure \ref{Fig1}a on the two-dimensional $(d_x,d_y)$ displacement space. The global energy minima are denoted by $Y^{\pm}$ and the saddle points by $S^{\pm}$ and $Y^0$.	
	(b) - (d) Schematic top-view of three LaSb$_2$ building blocks with
	different stacking sequences. The cell with label 1 denotes the bottom block. 
	(b) Sm-type orthorhombic structure with alternating stacking sites of $[S^{+}S^{-}]$; 
	(c) Yb-type orthorhombic structure with a sole stacking site of $[Y^0]$; and
	(d) Yb-like monoclinic structure with a sole stacking site of $[Y^{+}]$ resulting in a tilted out-of-plane lattice vector.
 (e) Total energies versus volume for the four crystal structures, where the solid curves are fits to the Birch-Murnaghan equation of state. 	
		}
	\label{fig:stacking}
\end{figure}

\begin{table*}
	\caption{\label{tab:bulk} 
	Calculated relative total energies, lattice constants, and angles of several LaSb$_2$ bulk phases corresponding to different stacking sequences. 	
	SG indicates the space group. The Sm-type structure is a base-centered orthorhombic cell and the lattice constants, $a$ and $b$,  are equivalent to the distance from the corner to the center of the conventional cell base.}
	\begin{ruledtabular}
		\begin{tabular}{l|ccr|cccccc}
			System & Stacking sequence & SG & $\Delta E$ (meV/La) & $a$ (\AA) & $b$ (\AA) & $c$ (\AA) & $\alpha$ ($^\circ$) & $\beta$ ($^\circ$) & $\gamma$ ($^\circ$) \\
			\hline
			Sm-type\cite{footnote} & $\cdots S^+S^-S^+S^-[S^+S^-] \cdots$ & 64 & 29 & 4.470 & 4.470 & 18.628 & 90.0 & 90.0 & 91.1 \\
			Yb-type & $\cdots Y^0Y^0Y^0Y^0Y^0[Y^0] \cdots$ & 63 & 45 & 4.477 & 4.546 & 17.607 & 90.0 & 90.0 & 90.0 \\
			Yb-alt  & $\cdots Y^+Y^-Y^+Y^-[Y^+Y^-] \cdots$ & 58 &  4 & 4.555 & 4.480 & 17.717 & 90.0 & 90.0 & 90.0 \\
			Yb-mono & $\cdots Y^+Y^+Y^+Y^+Y^+[Y^+] \cdots$ & 12 &  0 & 4.562 & 4.483 & 17.711 & 90.0 & 86.3 & 90.0 \\
   
			\hline
			Bulk exp.\cite{footnote, LaSb2_bulk} & & 64 &  & 4.42 & 4.42 & 18.56 & 90.0 & 90.0 & 91.2 \\
			This work & & 12 &  & 4.52 & 4.43 & 17.62 & 90.0 & 85.96 & 90.0 \\
   
		\end{tabular}
	\end{ruledtabular}
\end{table*}

The stacking arrangements for these materials can be constructed from QL building blocks as shown in Figure \ref{Fig1}(a), where the upper block is shifted relative to the lower one by a displacement vector $\vec{d} = (d_xa_0, d_yb_0)$. This construction or interpretation is possible because of the relatively weak bond between neighboring La-Sb layers, substantiated by the lower cleavage energy (Figure \ref{suppfig:cleavage}) and the longer inter-block La-Sb bond length ($> 3.5$ \AA) compared to those of the intra-block bonds.
The energy landscape of the two LaSb$_2$ building blocks on the two-dimensional $(d_x,d_y)$ displacement space is shown in Figure \ref{fig:stacking}(a). This is obtained using a relaxed unit block slab where $a_0 =b_0= 4.50$ \AA$\,$ are the calculated equilibrium in-plane lattice constants, and shifting it relative to an adjacent slab, with subsequent relaxation of the atomic positions. The calculations reveal (i) four global energy minima, $Y^{\pm}$ and (ii) six saddle points, $Y^0$ and $S^{\pm}$. The symbols mark half of them and $C_4$ rotation covers the unmarked special stacking sites. It is evident that the vertical stacking $(d_x,d_y) = (0,0)$ is not favored.  The $S^{\pm}$ sites, $(d_x,d_y)=(\pm 0.25, \pm 0.25)$, correspond to the displacement of the Sm-type LaSb$_2$ structure (space group $No. 64, Cmca$) consisting of building blocks with stacking sequence of $\cdots S^+S^-S^+S^- \cdots$ (see also Table~\ref{tab:bulk}). The Yb-type structure (space group $No. 63, Cmcm$) can also be explained with the stacking sequences of $\cdots Y^0Y^0Y^0Y^0 \cdots$. The two displacements between three consecutive building blocks are illustrated in Figure \ref{fig:stacking}(b) and (c) for  the Sm- and Yb-type structures, where the successive lateral displacements are compensated and both Sm- and Yb-type structures preserve orthorhombicity. We note that the energy map preserves the $C_4$ rotational symmetry, inherited from the building block ($a_0=b_0$). The broken tetragonal symmetry of Sm- and Yb-type LaSb$_2$ is thus related to  the displaced stacking sites that breaks the $C_4$ symmetry of the building block. Both phases, however, do not contain the lowest energy stacking sites $Y^{\pm}$ according to these calculations. Figure \ref{fig:stacking}(d) shows one potential structure with repeating $Y^+$ stacking, $\cdots Y^+Y^+Y^+Y^+ \cdots$,  where the non-compensating displacement along the $\hat{x}$ direction leads to shear deformation of the out-of-plane lattice vector rendering the system monoclinic. To further validate this method, we also have applied it to the SmSb$_2$ and YbSb$_2$ structures and have found that in each case, the expected stacking arrangements are reproduced (Figure \ref{suppfig:energy-surfaces}).
 
Table~\ref{tab:bulk} lists the calculated relative total energies of the different stacking sequences, as well as experimentally obtained values for bulk crystals\cite{LaSb2_bulk} and our thin films. The calculations point to the monoclinic structure (Yb-mono) as the ground state, with the previously synthesized Sm-type or proposed Yb-type phases exhibiting higher total energies. We note however, that this is the energy at $T=0$, and therefore is distinct from the free energy at elevated temperatures which contains phononic, magnetic, and electronic entropic contributions. The novel Yb-mono phase is consistent with the MBE-grown LaSb$_2$ structure where all the calculated lattice constants and monoclinic angle $\beta$ agree well with the diffraction data within an error of 1\%. Combining diffraction, calculation and microscopy results, we can umambiguously confirm the novel Yb-mono polymorph has been stabilized. 

Interestingly, the calculations also reveal a new orthorhombic phase (Yb-alt in Table~\ref{tab:bulk}, space group $No. 58, Pnnm$) with alternating stacking sequences of $\cdots Y^+Y^-Y^+Y^- \cdots$,  which is close in energy to the ground state Yb-mono structure.
In order to study the relative stability of the four crystal structures under hydrostatic pressure, we show in Figure \ref{fig:stacking}(e) the calculated total energies as a function of volume. We find that the Yb-mono structure remains the ground state under hydrostatic pressure, which on the other hand, induces a structural phase transition from the Sm-type to the Yb-type structures at 0.749 GPa, which is close to the calculated value in Ref~[\onlinecite{weinberger:2023}] of $\simeq$ 0.5 GPa.

\begin{figure}[ht]
	\includegraphics[width = 82mm]{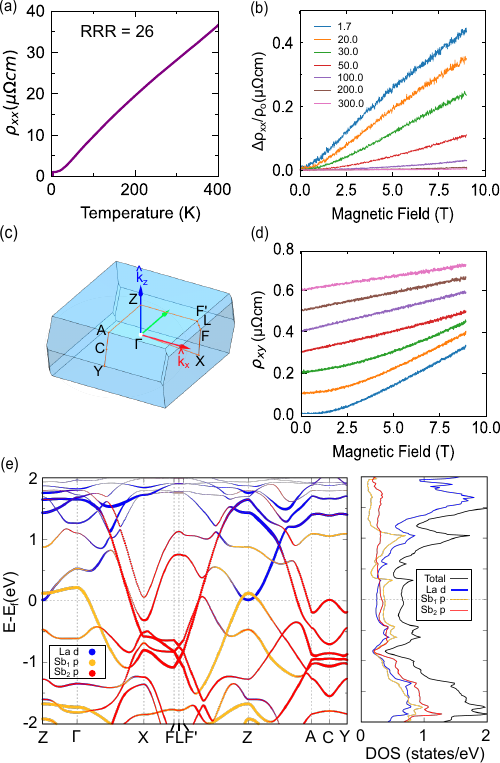}
	\caption{(a) Temperature-dependent longitudinal resistance. (b) Longitudinal magnetoresistance at different temperatures. (c) Brillouin zone and relevant high symmetry points of the monoclinic structure. (d) Hall resistance curves at different temperatures, offset for clarity. (e) Calculated band structure with color coded dispersion features associated with projections on various atomic sites, along with the corresponding DOS 
 (total DOS scaled by $\frac{1}{4}$).}
	\label{FigTransport}
\end{figure}

In Figure \ref{FigTransport}a we present temperature dependence of resistivity for a film of thickness $t=144$~nm. The films show metallic $\rho(T)$ with an RRR (R$_\mathrm{300K}$/R$_\mathrm{2K}$) of 26. No features with sharp changes in $d\rho/dT$ are observed up to $T$=400~K which would indicate the presence of a CDW transition or first-order structural transition. Comparison of the data during both warming and cooling shows slight increase in sample resistivity and slightly degraded RRR (Figure \ref{suppfig:warm-cool}) but no other anomalies. 
In Figure \ref{FigTransport}b and \ref{FigTransport}d, the symmeterized magnetoresistance (MR) and anti-symmeterized Hall resistivity  up to fields of $B$=9~T for several temperatures are given. 
The clear curvature of the Hall resistance indicates multiple carrier types while the MR displays linear behavior at high fields at low temperature, similar to bulk crystal data both in ambient conditions and under applied pressure, albeit with a reduced magnitude compared to bulk crystals.\cite{young:2003, Budko:1998} It was observed that while the behavior of the Hall resistivity was reasonably captured by a simple Drude two carrier model using the formulas in Ref.\cite{two_carrier}, the magnetoresistance could only be fit by this model for higher temperature scans, indicating that scattering mechanisms beyond the simple Drude picture are relevant as temperature is reduced (Figure \ref{fig:MR-SI}). 

In Figure \ref{FigTransport}c we plot the Brillouin zone of the monoclinic crystal structure with the relevant high symmetry points. The band structure along the high symmetry directions (orange path in (c)) and the corresponding orbital-projected density of states (DOS) are displayed in Figure \ref{FigTransport}e, where the blue, orange, and red colors denote projections on the  La-5$d$, Sb$_1$/5$p$ and Sb$_2$/5$p$ orbitals (Sb$_1$ are on the La/Sb corrugated layer and Sb$_2$ are on the 2D Sb sheets in see Figure \ref{Fig1}(a)). 
The relatively flat bands along the Z-$\Gamma$, X-F and A-C-Y directions indicate lower carrier mobility along the z-direction, in contrast to the more dispersive Sb$_1$/5p and Sb$_2$/5p bands crossing the Fermi level along the $\Gamma$-X and F’-Z directions, indicating higher carrier mobility along the in-plane directions similar to those in the bulk orthorhombic LaSb$_2$ structure\cite{ruszala,qiao:2022,palacio:2023} and in other topological square-net materials.\cite{klemenz:2019} The presence of both hole-like and electron-like pockets at the Fermi level is in agreement with the observation of multiple carrier types from the Hall measurements.

\begin{figure}[ht]
	\includegraphics[width = 82mm]{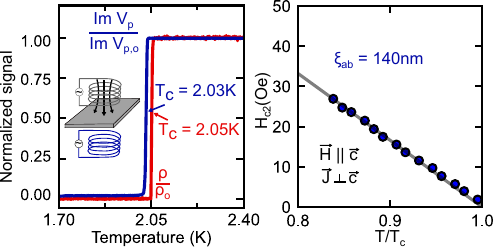}
	\caption{(a) Superconductivity measured in a 144~nm film via resistance (red) and mutual inductance (blue) methods. (b) Critical field as a function of reduced temperature, yielding an estimate of the in-plane coherence length.}
	\label{FigSuperconductivity}
\end{figure}

In Figure \ref{FigSuperconductivity}a we present data showing the presence of a sharp superconducting transition with resistive \Tc~$\approx$~2.05~K. The films display zero resistance and clear flux expulsion, as evidenced by a drop in voltage on the secondary coil in a mutual inductance experimental geometry (inset of \ref{FigSuperconductivity}a). The measured T$_c$ in the two experiments agree within 1\%. In Figure \ref{FigSuperconductivity}b, the critical magnetic field $H_\mathrm{c2}$ for fields applied along the $c$-axis with current applied parallel to the $a-b$ plane is plotted. Determination of Hc2 was performed by fitting the transition isotherms to a sigmoid function $\rho=\frac{1}{1 + exp(-k(H-H_{c2}))}$ where k, A and $H_{c2}$ were fitting parameters. This fit implies that $H_{c2}$ is given by the intersection with half the normal state resistance. The full set of critical field isotherms and fits is given in \ref{suppfig:Critical_field}. Data were taken every 25mK between 2.05K and 1.71K. In \ref{FigSuperconductivity}b, we fit the critical field data using a single-band, Ginzburg-Landau model\cite{tinkham2004introduction} to obtain estimates of the in-plane coherence length $\xi_{ab} = 140$~nm. This value is also in agreement with the findings in bulk crystals under pressure.\cite{guo:2011}

One possibility for the stabilization of the monoclinic LaSb$_2$ is related to epitaxial strain or other perturbations induced by the substrate. The square symmetry of the MgO substrate would tend to favor layers with reduced orthorhombicity and lattice mismatch with the substrate. In the monoclinic phase, the $a$ and $b$ axes differ by $\approx$1.8\% compared to approximately $\approx$2.5\% for the ambient phase structure. When comparing the lattice mismatch between the two structures and the substrate, it would appear that the orthorhombic SmSb$_2$ \cite{footnote2}, with mismatches $\approx$6\% and $\approx$3\% for the $a$ and $b$ axes, would be favored over the monoclinic phase which shows mismatches of $\approx$7\% and $\approx$5\%. However in both cases, these lattices mismatches are so large, that is it unlikely that coherent strain persists in the epilayer beyond the first 1-2~QL. Taken together, these findings indicate that while layer-by-layer growth stabilizes the monoclinic LaSb$_2$, this observation is unlikely to be the result of epitaxial strain. Another possibility for the observation of the monoclinic phase in thin films, yet not in the bulk crystals, is the lower growth temperature used in the MBE growth processes ($\approx$500$^o$C $< T_\mathrm{melt}$ = 630$^\circ$C).\cite{Okamoto2019} During flux growth, crystals are obtained by slowly cooling the melt from the liquidus temperature ($\approx$1000$^o$C), over a period of tens of hours and decanting the liquid flux slightly above the solidification temperature of the desired phase.\cite{flux_growth} Given the results of the 0~K DFT simulations, monoclinic LaSb$_2$ may therefore form a low-temperature crystalline phase that would be preferentially stabilized by the vapor-solid reaction of MBE.

We finally note that as of the time of writing this report, we are unaware of single-crystal diffraction data reported for bulk-LaSb$_2$ at high $P$. The micacity of crystals makes such experiments challenging. Previous studies\cite{weinberger:2023} have proposed YbSb$_2$ or EuSb$_2$ as candidate structures for the high-$P$ phase on the basis of DFT calculations. Our calculations agree that while the YbSb$_2$ structure is more stable compared to the SmSb$_2$ structure under pressure, the monoclinic $Y^+$ structure is the ground state. We therefore propose that the application of pressure results in a structural transition to either the $Y^+$ polymorph or perhaps a similar orthorhombic structure labelled $[Y^+Y^-]$ which is nearly degenerate. The association of the $Y^+$ phase with the high-pressure phase of the bulk crystal is further bolstered by the qualitative features of transport and superconducting behavior, namely, the absence of resistance anomalies at high temperature and a sharp superconducting transition at 2K. In bulk LaSb$_2$, the superconducting transition is broad, with onset temperatures of about $T$=2~K, yet only reaching a fully superconducting state at $T$=0.4~$\sim$0.5K\cite{guo:2011}. The transition sharpens with $P$ with concomitant suppression of the high temperature resistance anomaly.\cite{weinberger:2023, Budko:2023} A maximum T$_c \approx$~2~K was observed under applied pressure of 4.5kbar at which point the anomaly is completely absent.\cite{guo:2011,weinberger:2023} We note that the strongly 2D-like dispersion in the other square-net systems such as SmSb$_2$-type LaSb$_2$ is significantly different in our structure (Figure \ref{suppfig:fermi} for discussion), which may play a role in the suppression of the resistance anomaly associated with a structural transition that competes with superconductivity. Further investigations into the superconducting phase will form the basis of future studies. 

In summary, we have experimentally and theoretically presented a newly stable polymorph of in the family of rare-earth diantimonides in thin films of LaSb$_2$ deposited on MgO (001). The films are superconducting with a critical temperature enhanced relative to reports on bulk crystals under ambient conditions. The coherence length of the superconductor is comparable to the film thickness, pointing towards this family of materials as a promising platform to study superconductivity with varying lattice motifs and layering orders which may be inaccessible or difficult to characterize in bulk samples.

\section*{Acknowledgements}
We thank Paul Canfield and Sergey Budko for helpful conversations and critical reading of the manuscript. J.F acknowledges funding provided by the the Air Force Office of Scientific Research (Grant number FA9550-22-1-0463), the Gordon and Betty Moore Foundation’s EPiQS Initiative (Grant number GBMF10638), and the Institute for Quantum Information and Matter, an NSF Physics Frontiers Center (NSF Grant PHY-1733907). The work at CSUN acknowledges support from the NSF-PREP CSUN/Caltech-IQIM Partnership (Grant number 2216774) and NSF-PREM (Grant number DMR-1828019). We acknowledge the Beckman Institute for their support of the X-Ray Crystallography Facility at Caltech. Electron microscopy was performed at the Center for Electron Microscopy and Analysis (CEMAS) at The Ohio State University.

\section*{Conflict of Interest}
The authors declare no conflict of interests.

\section*{Data Availability Statement}
The  data  that  support  the  findings  of  this  study  are  available  from  the  corresponding author upon reasonable request.

\clearpage

\section*{}
\bibliography{bib}

\begin{thebibliography}{47}
\bibitem[{foo({\natexlab{a}})}]{footnote}
\bibinfo{note}{To facilitate comparison with the monoclinic phase, the Niggli reduced cell of the SmSb$_2$ ambient phase is used. This unit cell has non-orthogonal a,b lattice constants of equal length which is approximately $\sqrt{2}$ smaller than the lengths of the conventional axes.}
\bibitem[{foo({\natexlab{b}})}]{footnote2}
\bibinfo{note}{Comparisons here are made to the conventional unit cell for the ambient phase SmSb$_2$-type LaSb$_2$}.
\end{thebibliography}

\clearpage

\renewcommand{\thefigure}{S\arabic{figure}}

\setcounter{figure}{0}

\subsection*{Supplementary Data}

\section{Experimental Section}
\begin{figure*}[ht]
	\centering
	\includegraphics[width=7.5cm]{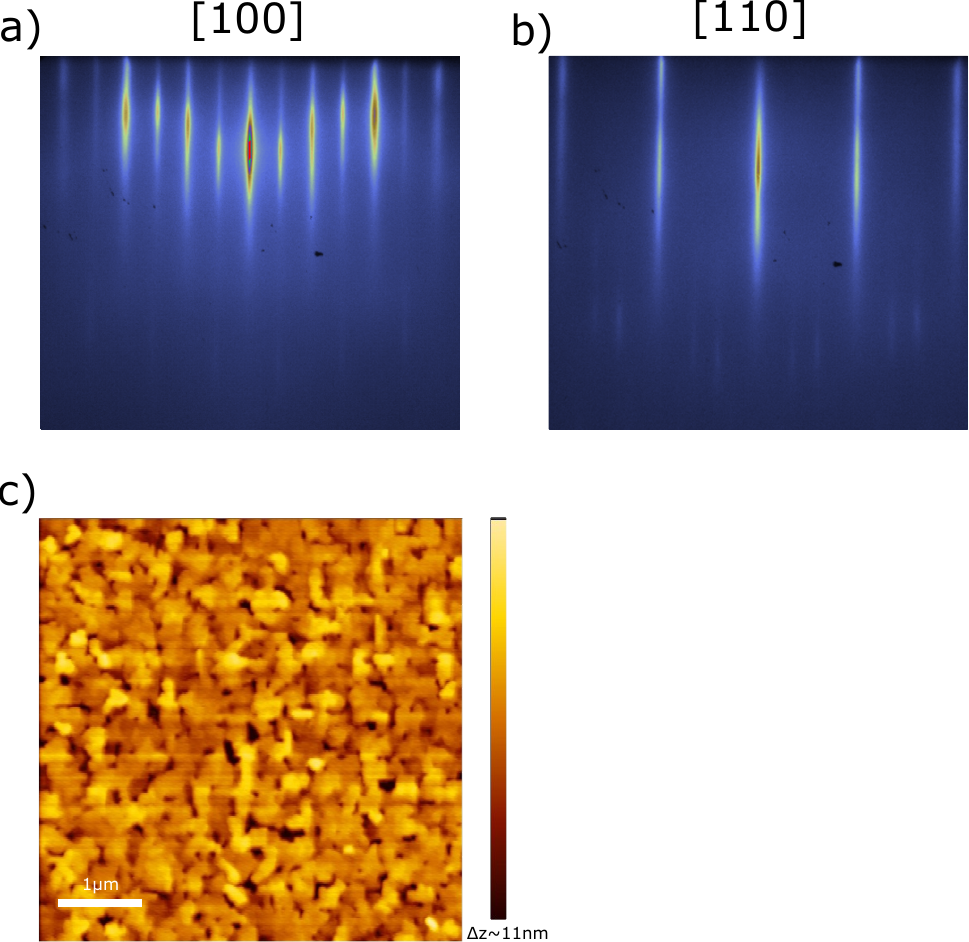}
	\caption{a) RHEED taken at the end of grown along [100] b) [110]. c) AFM of surface of 24nm film.}
	\label{suppfig:AFM-RHEED}
\end{figure*}

\begin{figure*}[ht]
	\centering
	\includegraphics[width=15cm]{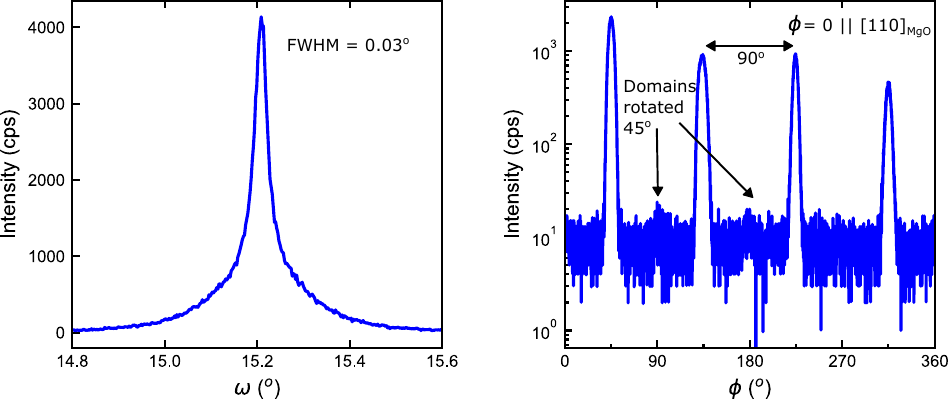}
	\caption{a) Rocking curve of (003) peak showing FWHM of 0.03$^{\circ}$ b) Phi-scan demonstrating the epitaxial relationship described in the main text. Minor domains rotated 45$^{\circ}$ can be seen.}
	\label{suppfig:RC-Phi}
\end{figure*}

\begin{figure*}[ht]
	\centering
	\includegraphics[width=15cm]{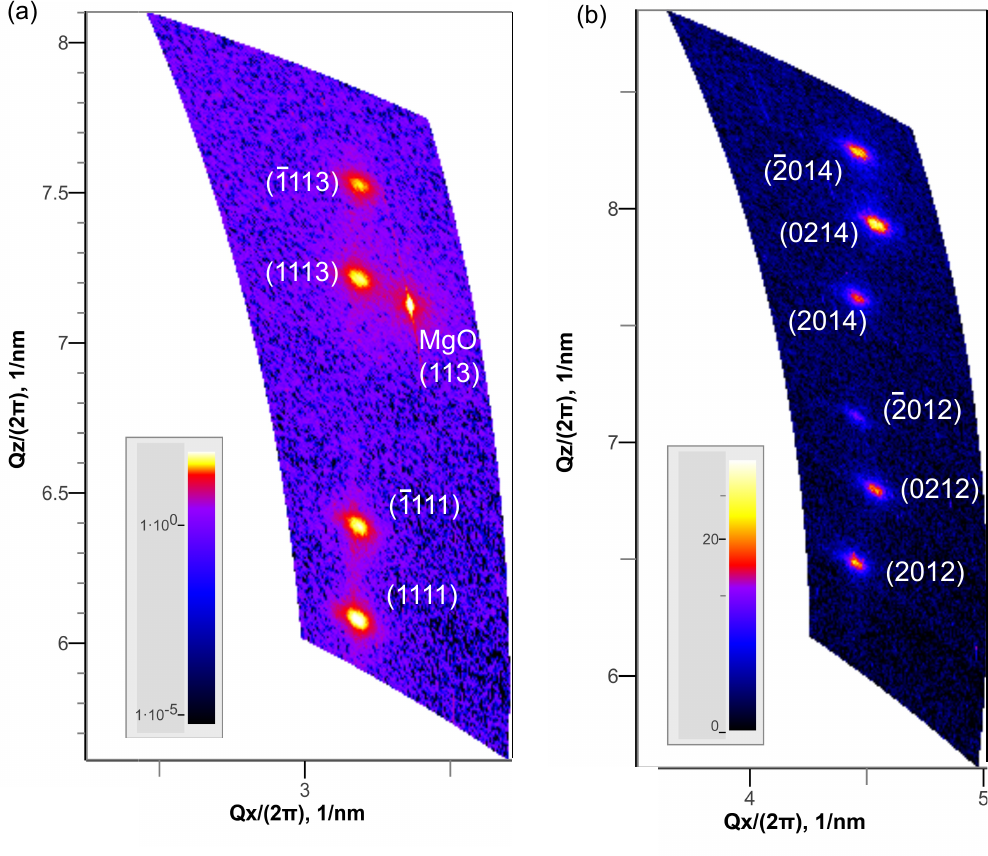}
	\caption{Additional reciprocal space maps used to verify the extinction conditions and determine the space group of the grown film. The above maps and those in the main text imply k+l = 2n for general h,k,l; l = 2n for h, k = 0, l; k+l = 2n for h = 0, k, l. Comparing with the listings in the international tables (International Tables for Crystallography (2016). Vol. A, Section 1.6.5, pp. 114–128.), these conditions, along with the identification as a monoclinic crystal imply 3 possible space groups: A2, Am, A2/m. Of these, A2/m is determined on the basis of TEM measurements which confirm the atomic arrangement predicted by DFT calculations.}
	\label{suppfig:RSM-SI}
\end{figure*}

\begin{figure*}
	\centering
	\includegraphics[width=15cm]{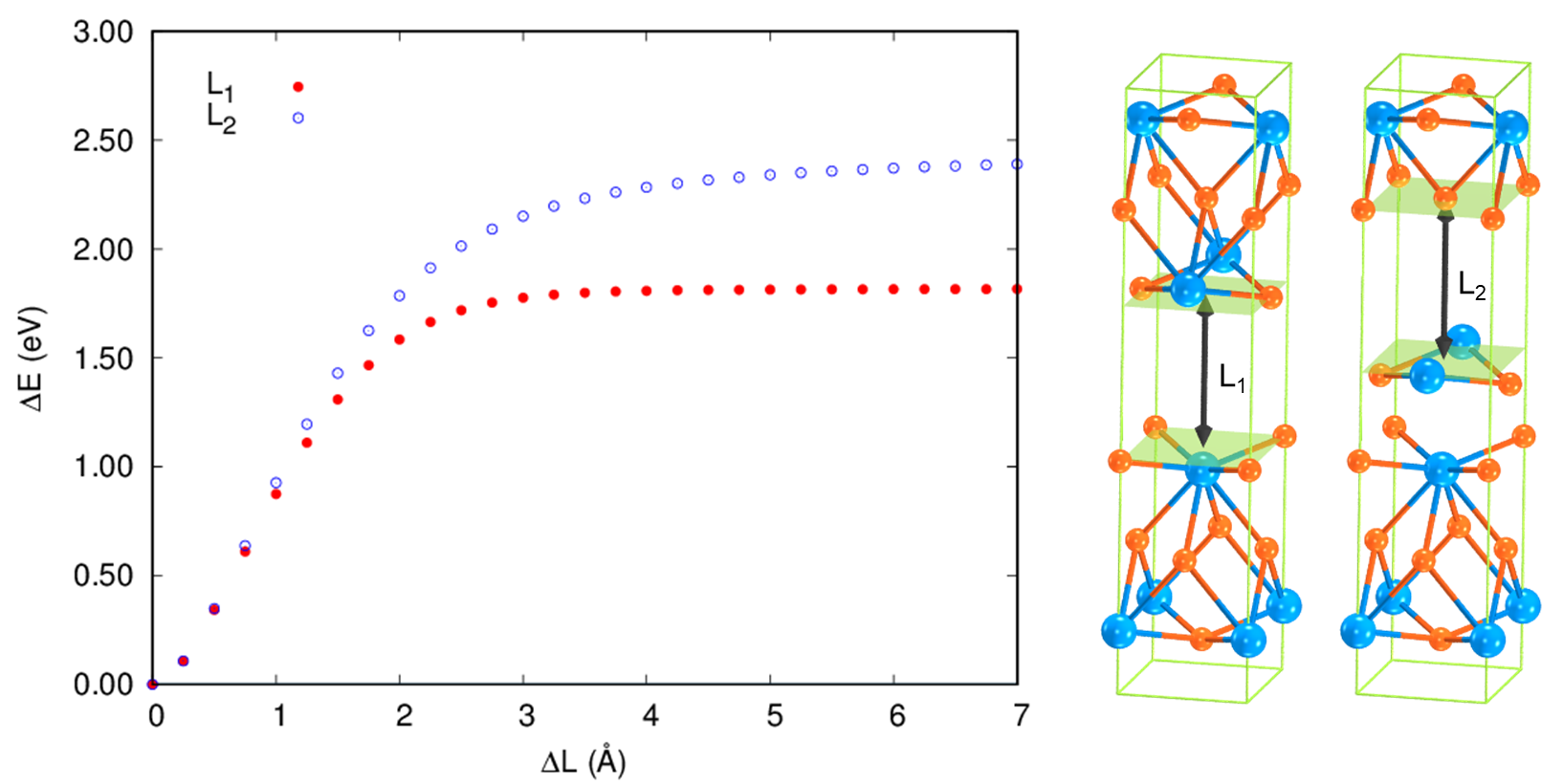}
	\caption{Calculated total energy difference, $\Delta$E, versus cleavage separation, $\Delta$L, between  two LaSb$_2$ unit building blocks in the Y$^{\pm}$ ground-state configuration [Figure \ref{fig:stacking}(a)]  for separation between (a) two successive La-Sb corrugated  layers (red circles)  shown in the left atomic structure and (b)  successive La-Sb corrugated layer and the square Sb sheet (blue circles), shown in the right atomic structure, respectively. $\Delta$E converges to the ideal cleavage energy as the separation between the neighboring building blocks increases. The substantially lower cleavage energy for two successive La-Sb corrugated layers compared to that for the La-Sb corrugated layer and the square Sb sheet, clearly demonstrates that the interlayer bonding between two successive La-Sb corrugated layer is weaker, thus allowing their easier glide.}
	\label{suppfig:cleavage}
\end{figure*}

\begin{figure*}
	\centering
	\includegraphics[width=17cm]{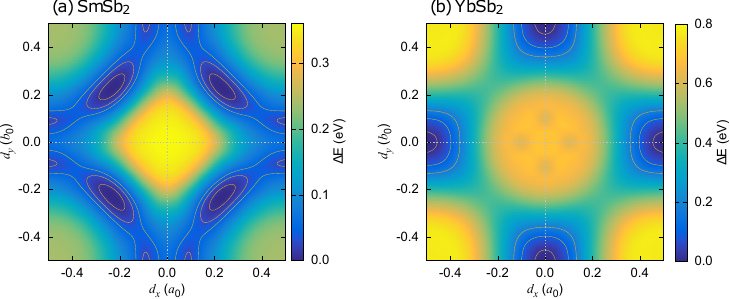}
	\caption{Energy surfaces for the SmSb$_2$ structure and YbSb$_2$ structure generated in the same manner as that used in the main text. Results indicate that the expected structures are the most stable for these compounds.}
	\label{suppfig:energy-surfaces}
\end{figure*}

\begin{figure*}
	\centering
	\includegraphics[width=11cm]{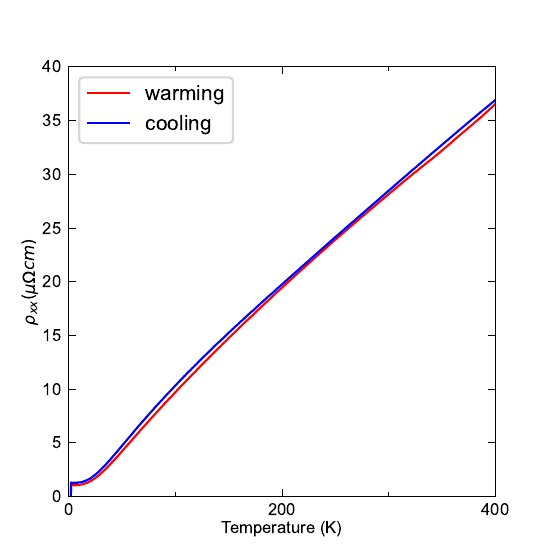}
	\caption{Comparison of warming and cooling curves up to 400~K. Data show no anomalies associated with CDW or other structural transition. }
	\label{suppfig:warm-cool}
\end{figure*}

\begin{figure*}
	\centering
	\includegraphics[width=17cm]{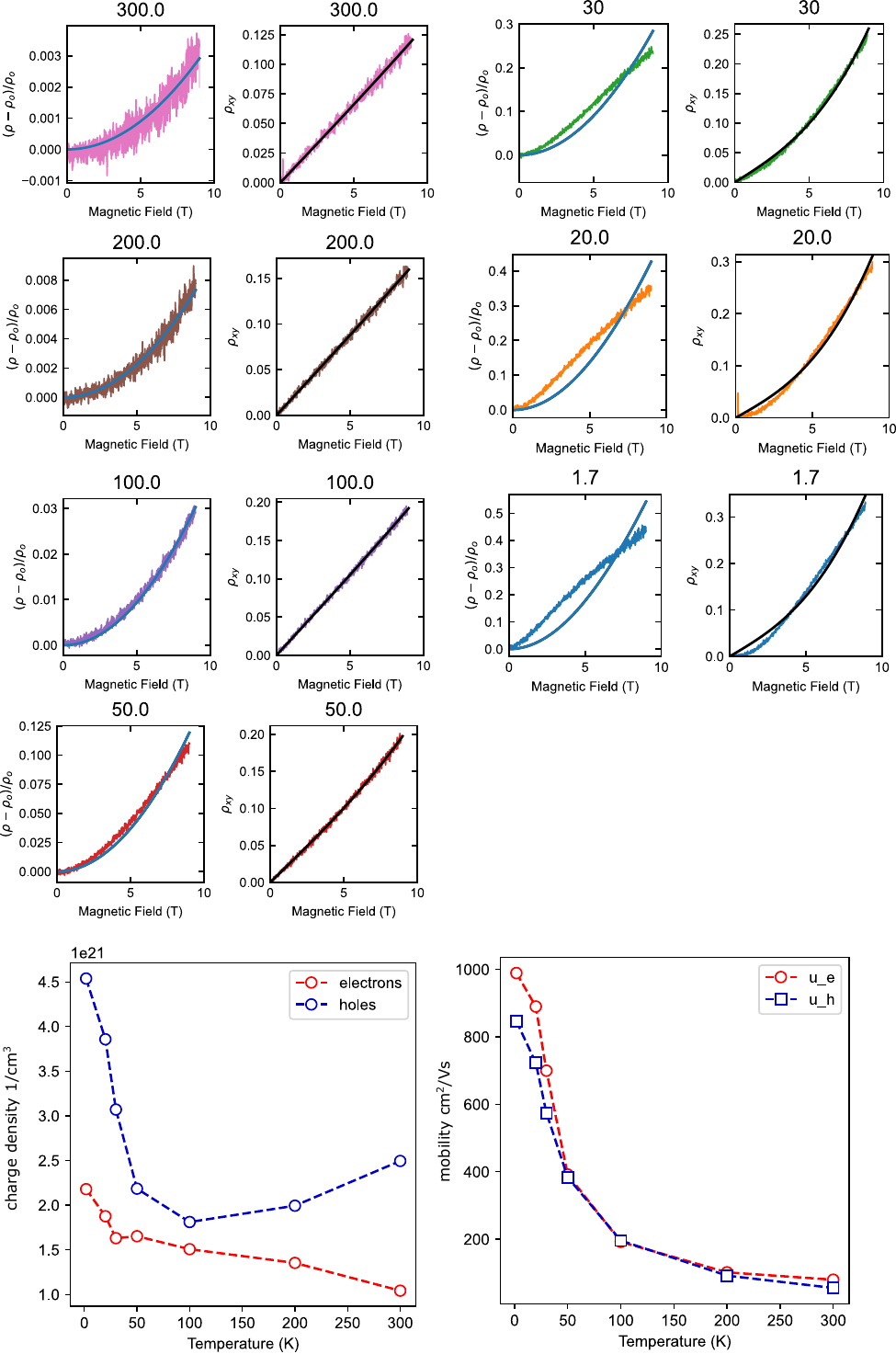}
	\caption{(a) fits to two-carrier Drude model for increasing temperature. $\rho_{xx} = \frac{1}{e}\frac{(n_h\mu_h + n_e\mu_e) + (n_h\mu_e + n_e\mu_h)\mu_h\mu_eB^2}{(n_h\mu_h + n_e\mu_e)^2 + (n_h-n_e)^2\mu_h^2\mu_e^2B^2}$, $\rho_{xy} = \frac{B}{e}\frac{(n_h\mu_h^2 - n_e\mu_e^2) + (n_h-n_e)\mu_h^2\mu_e^2B^2}{(n_h\mu_h + n_e\mu_e)^2 + (n_h-n_e)^2\mu_h^2\mu_e^2B^2}$. (b), (c) variation with temperature of charge density and mobility, respectively}
	\label{fig:MR-SI}
\end{figure*}

\begin{figure*}
	\centering
	\includegraphics[width=11cm]{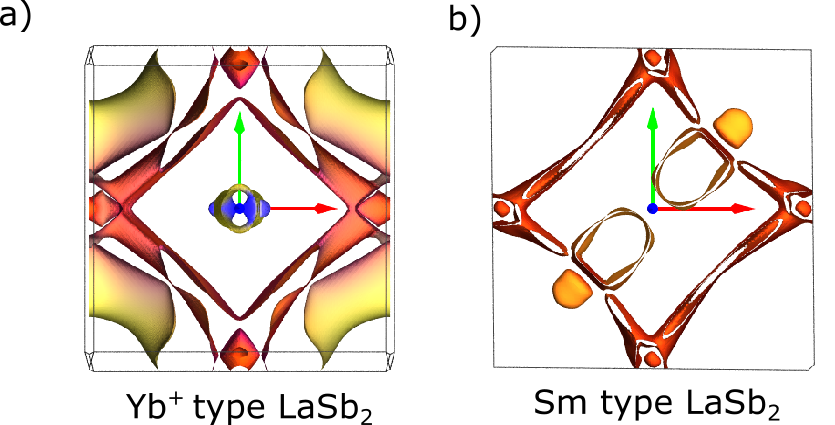}
	\caption{Fermi surface (FS) projections for (a) thin film Yb$^+$-type LaSb$_2$ and (b) bulk Sm-type LaSb$_2$. Suppression of a competing CDW is thought to lead to enhancement of superconductivity in monoclinic LaSb$_2$. In this system, there is strong hybridization with Sb$_1$-p which introduces band dispersion along the $k_z$ direction and breaks 2D like features. Overall, the FS of the monoclinic structure is distinct from the highly nested FS sheets of the orthorhombic Sm-type structure and could be a possible origin of CDW suppression and enhancement of the superconducting Tc. }
	\label{suppfig:fermi}
\end{figure*}

\textit{Molecular beam epitaxy}: Films were prepared using a molecular beam epitaxy (MBE) apparatus at a base pressure of $\approx 10^{-10}$ Torr. Films were grown upon MgO (001) ($a$=0.421~nm) substrates which were thermally prepared using a home-built CO$_2$ laser heating apparatus,\cite{llanos2023supercell} and then later transferred to the growth chamber equipped with a conventional resistive SiC coil. La and Sb were both provided from conventional effusion cells. All growths begin with a low temperature buffer layer with La and Sb codeposited at beam flux pressures of 6.5$\times10^{-9}$ mbar and 1.5$\times10^{-7}$ mbar, at growth temperature 315$^\circ$C. The buffer layer is annealed at 615$^\circ$C for 5 minutes before being cooled to 520$^\circ$C, where the rest of the film is grown with La and Sb codepositing at beam flux pressures of 1$\times10^{-8}$ mbar and 1.5$\times10^{-7}$ mbar. Growth rate was estimated using Laue fringes observed in lower thickness films and was found to be 0.033 \AA/s. To reduce degradation in air, all films were capped with amorphous Ge \textit{in-situ} at room temperature.  Reflected high energy electron diffraction (RHEED) was obtained at an accelerating voltage of 20~kV. RHEED and AFM images from a 24nm film can be found in \ref{suppfig:AFM-RHEED}. RHEED patterns are streaky high-contrast indicating good crystallinity. Images also reveal a 2x1 surface reconstruction. AFM was used to asses the quality of the surface where it can be seen that the surface consists of large, flat regions punctuated by deeper cavities.  

\textit{X-ray diffraction}: X-ray diffraction data was obtained using a Rigaku Smartlab diffractometer using a 2-bounce Ge (220) monochromator and Cu-$K\alpha1$ radiation. 

\textit{Transmission electron microscopy}:
Scanning Transmission Electron Microscopy (STEM) measurements were performed on a cross-sectional lamella prepared through focused ion beam (FIB) lift-out using a FEI Helios NanoLab 600 DualBeam. The samples underwent thinning with 5~kV Ga ions, followed by a final polish at 2~kV to minimize surface damage. The imaging was conducted using the Thermo Scientific Themis Z S/TEM operating at 300~kV, featuring a semi-convergence angle of 30~mrad. A High-Angle Annular Dark-Field (HAADF) detector with an angular range of 64-200~mrad was utilized to capture HAADF-STEM images. This imaging process involved acquiring a series of 20 images (200 ns per frame), which were subsequently averaged to produce images with a high signal-to-noise ratio. STEM- Energy dispersive X-ray spectroscopy (EDX) data is acquired using a Super-X EDX detector.

\textit{First principles calculations}:
The first principles calculations are performed by using the Vienna $ab$ $initio$ simulation package.
The wave functions and pseudopotentials are treated with the projected augmented wave method. 
The exchange-correlation functional is described with Perdew-Burke-Erenzhoff approach. The plane wave
basis cut-off energy is set to 500 eV and the momentum space is sampled with $\Gamma$-centered $16 \times 16 \times 1$ and $16 \times 16 \times 6$ meshes for the slab and bulk calculations, respectively. All structures are relaxed within energy convergence criteria of $10^{-5}$  eV. Spin-orbit coupling is not included.
The single building block is fully relaxed first with vacuum layer thicker than 15 \AA. The energy landscape upon displacement is then calculated with a sampling width of $\Delta d_{x,y} = 0.05$ and interpolated by using spline method. The in-plane lattice constants of the building block and the displaced in-plane coordinates of ions are fixed and only out-of-plane coordinates are relaxed.

\textit{Transport Measurements}:
Electrical data was obtained using four-point van der Pauw geometry measurements in a Quantum Design Dynacool system. Samples were cut in approximately square shape and sheet resistance was obtained by numerically solving the van der Pauw equation. For superconducting measurements, $T_c$ was determined to be the temperature at which the sheet resistance crosses half the value of the normal state sheet resistance by fitting a sigmoid function ($\rho=\frac{1}{1 + exp(-k(H-H_{c2}))})$ to the transition. 

\textit{Mutual Inductance Measurements}:
The LaSb${}_2$ film was held upright on the cryostat puck via a 3D printed housing. Two solenoids (5 mm in length, 2.5 mm in diameter) were inserted into the housing centered co-axially on either side of the film, and the sample was thermally and electrically grounded by a gold wire soldered to the corner of the film. An AC driving current (50 $\mu$A at 10.13 kHz) was passed through one coil using an SR830 lock-in amplifier while the first-harmonic pickup voltage was measured in the second coil with the same lock-in (300 ms time constant). An out-of-plane magnetic field of 10 Oe was applied to compensate for trapped flux in the cryostat system's magnet. While sweeping the sample's temperature at 10 mK/minute through its superconducting transition, the onset of the Meissner effect was observed as a substantial drop in the induced voltage across the pickup coil, indicating screening of the drive coil's magnetic field by the superconducting film.
\newpage


\begin{figure*}[ht]
	\centering
	\includegraphics[width=11cm]{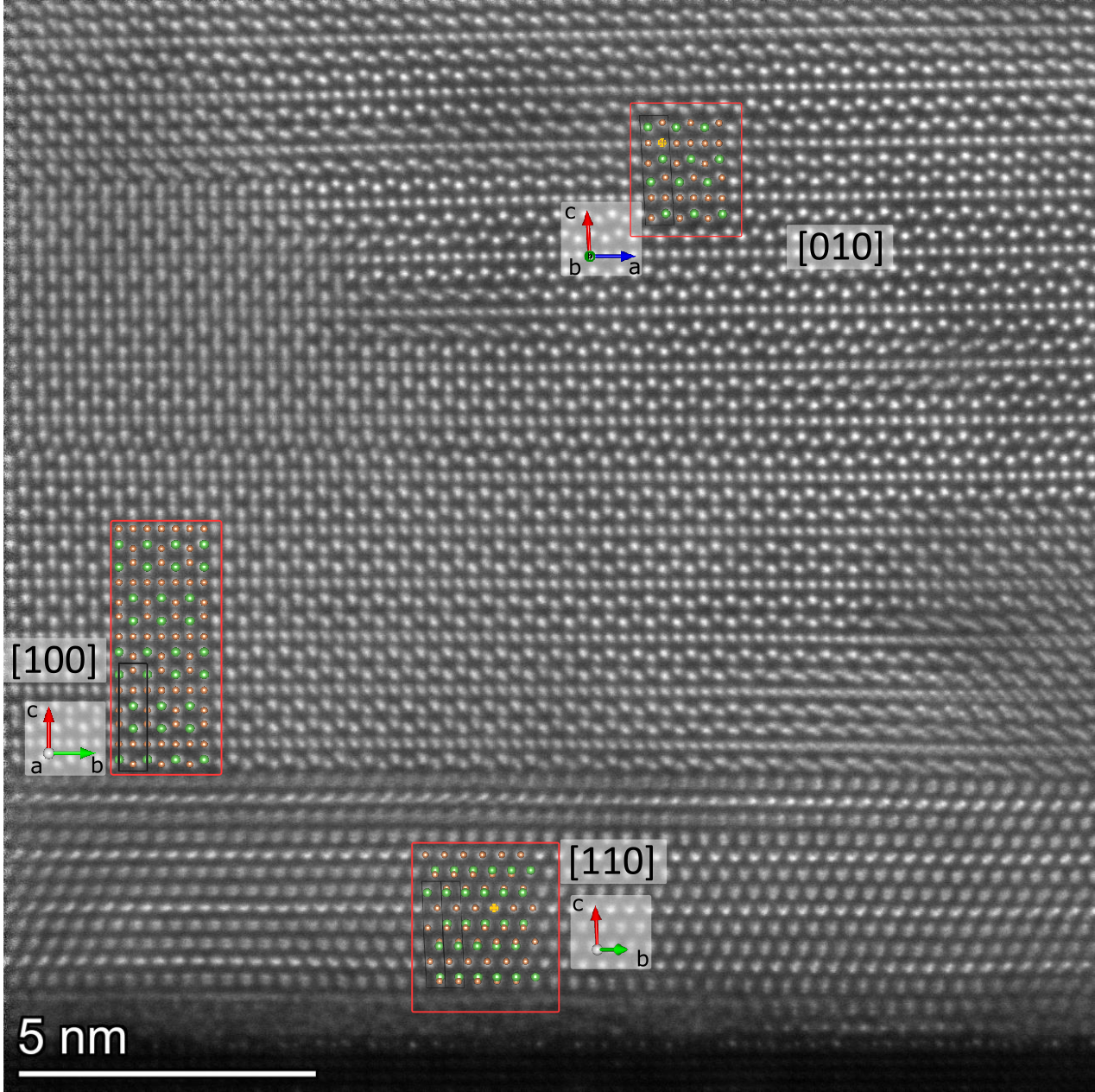}
	\caption{Wide area TEM showing the presence of a-b axis twinning and rotated domains oriented parallel to [110], close to the interface.}
	\label{suppfig:TEM-domains}
\end{figure*}
\begin{figure*}[ht]
	\centering
	\includegraphics[width=11cm]{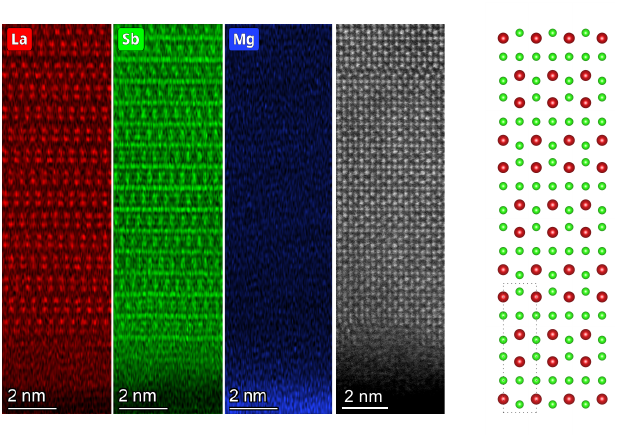}
	\caption{Elemental mapping of films, including the monoclinic structure for reference. The survey HAADF and EDX elemental maps in shows the monoclinic (C2/m) LaSb2 structure with zig-zag of Sb and La and Sb sheets. The La/Sb ratio with absorption and ionization cross-section correction for the foil with a thickness of 30 nm is 0.566. The structure with Sb and La atoms all aligned along [100] direction is unique to the monoclinic structure. MgO is beam sensitive structure and the structure is damaged while collecting EDX data. The EDX elemental maps acquired from a wide field of view shows uniform distribution of La and Sb with similar La/Sb ratio for different regions. }
	\label{suppfig:EDX}
\end{figure*}

\begin{figure*}[ht]
	\centering
	\includegraphics[width=11cm]{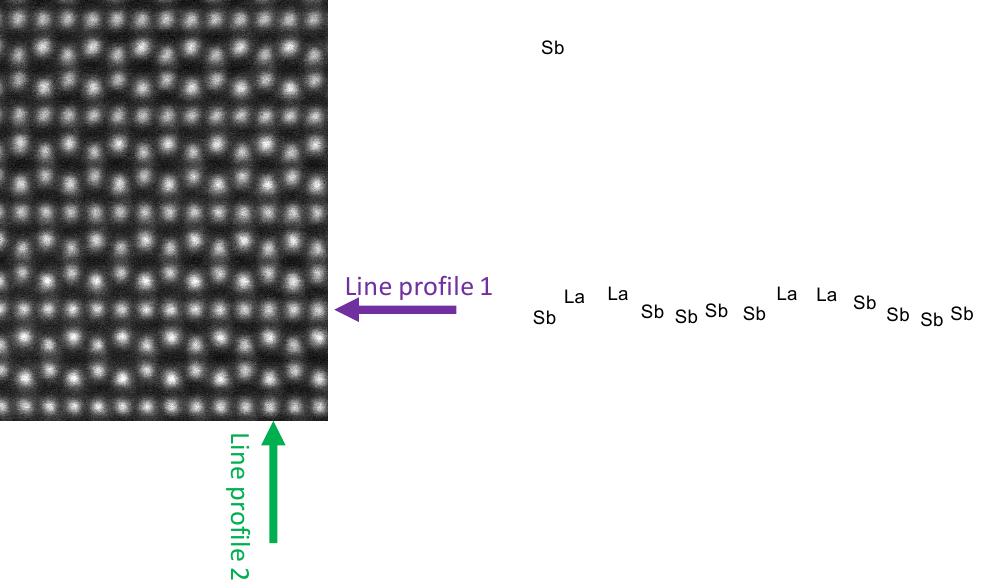}
	\caption{Line profiles on atomically resolved images along the [100] zone axis. The uniform ampltide across the Sb sheet indicates an absence of distortion along the c-axis direction, in contrast to the YbSb$_2$ structure}
	\label{suppfig:linecut}
\end{figure*}

\begin{figure*}[ht]
	\centering
	\includegraphics[width=13cm]{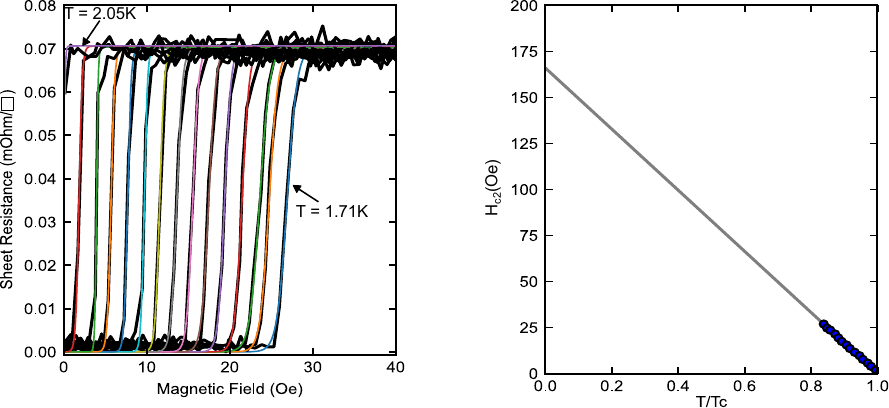}
	\caption{a) Critical field isotherms showing the increase in Hc2 vs temperature. Estimation of Hc2 was performed by fitting the transition to a sigmoid function $\rho=\frac{1}{1 + exp(-k(H-H_{c2}))}$ where k, A and $H_{c2}$ were fitting parameters. Fits are displayed along with raw data. b) extrapolation of the Ginzburg-Landau fit to zero temperature}
	\label{suppfig:Critical_field}
\end{figure*}
\end{document}